\documentclass[a4paper]{jpconf}
\usepackage{graphicx}
\begin{document}
\title{Effects of magnetic anisotropy on spin dynamics of ferromagnetic frustrated chain}

\author{Hiroaki Onishi}

\address{Advanced Science Research Center,
Japan Atomic Energy Agency,
Tokai, Ibaraki 319-1195, Japan
}

\ead{onishi.hiroaki@jaea.go.jp}

\begin{abstract}
By exploiting density-matrix renormalization group techniques,
we investigate the spin dynamics of a spin-1/2 one-dimensional $J_{1}$-$J_{2}$ XXZ model
with competing ferromagnetic $J_{1}$ and antiferromagnetic $J_{2}$ exchange couplings
under applied magnetic fields.
Numerical results of spin excitation spectra show that
in the field-induced spin quadrupole regime,
the longitudinal component has a gapless mode
and the transverse component has a gapped mode
irrespective of the exchange anisotropy.
The excitation gap of the transverse spin excitation increases
as the exchange anisotropy increases
over the XY-like and Ising-like regions,
demonstrating that two-magnon bound states are stabilized due to the easy-axis anisotropy.
\end{abstract}

%-----------------------------------------------------------
% Introduction
%-----------------------------------------------------------
\section{Introduction}

Frustrated quantum spin systems in low dimensions have provided a fascinating playground
to explore strange new phases that have no classical analogue,
and rich physics of quantum phase transitions between them.
One of the simplest models with frustration and low-dimensionality is
a spin-1/2 one-dimensional $J_{1}$-$J_{2}$ Heisenberg model,
in which the nearest-neighbor exchange interaction $J_{1}$
and the next-nearest-neighbor exchange interaction $J_{2}$ compete with each other.
Note that when $J_{2}$ is antiferromagnetic,
frustration occurs in either case of antiferromagnetic or ferromagnetic $J_{1}$.
Recent studies on the ferromagnetic $J_{1}$ case in a magnetic field have revealed that
the ground state is a vector chiral (VC) state in low magnetic fields,
while it turns to an exotic Tomonaga-Luttinger-liquid state
in high magnetic fields~\cite{Vekua2007,Hikihara2008,Sudan2009,Sato2009}.
In the high-field phase,
the formation of two-magnon bound states leads to a spin quadrupole state.
The longitudinal spin and quadrupole correlations are quasi-long-ranged,
while the transverse spin correlation is short-ranged.
The so-called spin nematic (SN) state is realized
in the vicinity of the saturation,
while the ground state shows a crossover to a spin-density-wave (SDW) state
with lowering the magnetic field.
The ground-state phase diagram in the presence of the exchange anisotropy
has also been analyzed numerically~\cite{Heidrich-Meisner2009}.
As for the experimental realization of the quadrupole state,
LiCuVO$_{4}$ has been studied
as a possible candidate material.
Neutron diffraction experiments have shown that
the field dependence of an incommensurate magnetic peak appearing in high magnetic fields
is well explained by theoretical predictions of the quadrupole state~\cite{Masuda2011}.
Inelastic neutron scattering measurements have been carried out at zero magnetic field,
indicating the existence of a two-spinon continuum~\cite{Enderle2010},
whereas the excitation spectrum in the high-field phase has not been reported yet.

To clarify the property of the field-induced quadrupole state
from the viewpoint of the spin dynamics,
we have studied the dynamical spin structure factor
of the spin-1/2 one-dimensional $J_{1}$-$J_{2}$ Heisenberg model
by exploiting numerical methods
such as a dynamical density-matrix renormalization group~\cite{Onishi2014}.
In this paper,
to gain further insight into the effect of the exchange anisotropy,
we analyze the $J_{1}$-$J_{2}$ XXZ model.
We present numerical results on the anisotropic behavior of
longitudinal and transverse spin excitation spectra,
and discuss the dependence on the exchange anisotropy.

%-----------------------------------------------------------
% Model and method
%-----------------------------------------------------------
\section{Model and method}

We consider a spin-1/2 $J_{1}$-$J_{2}$ XXZ model on a one-dimensional chain with $N$ sites,
described by
\begin{equation}
  H =
  J_{1} \sum_{i} ( S_{i}^{x}S_{i+1}^{x}+S_{i}^{y}S_{i+1}^{y}+\Delta S_{i}^{z}S_{i+1}^{z})
  +J_{2} \sum_{i} ( S_{i}^{x}S_{i+2}^{x}+S_{i}^{y}S_{i+2}^{y}+\Delta S_{i}^{z}S_{i+2}^{z})
  -h \sum_{i} S_{i}^{z},
\label{eq:H-xxz-hz}
\end{equation}
where
$S_{i}^{\alpha}$ is the $\alpha$ component of the spin-1/2 operator at site $i$,
$J_{1} \, (<0)$ and $J_{2} \, (>0)$ denote
the ferromagnetic nearest-neighbor and
the antiferromagnetic next-nearest-neighbor exchange interactions, respectively,
$\Delta$ is the anisotropy of the exchange interaction,
and $h$ is the magnetic field along the $z$ direction.
In the present study,
we fix $J_{1}=-1$ and $J_{2}=1$,
and examine the dependence on $\Delta$.

We investigate dynamical properties
by using density-matrix renormalization group (DMRG)
methods in open boundary conditions~\cite{White1992,Jeckelmann2002}.
To clarify the characteristics of anisotropic magnetic excitations,
we calculate the $z$ and $x$ components of the dynamical spin structure factor,
given by
\begin{equation}
  S^{\alpha}(q,\omega) =
  -\lim_{\eta \rightarrow 0}\frac{1}{\pi}{\rm Im}
  \langle \psi_{\rm G} \vert
  S_{q}^{\alpha \dagger}
  \frac{1}{\omega+E_{\rm G}-H+{\rm i}\eta}
  S_{q}^{\alpha}
  \vert \psi_{\rm G} \rangle,
\end{equation}
where $\vert \psi_{\rm G} \rangle$ is the ground state
and $E_{\rm G}$ is its eigenenergy.
We set an infinitesimal value $\eta$ to $0.1$.
The momentum representation of the spin operator in open boundary conditions is given by
\begin{equation}
  S_{q}^{\alpha} = \sqrt{\frac{2}{N+1}} \sum_{i} S_{i}^{\alpha} \sin (qi).
\end{equation}
For the calculation of the dynamical spin structure factor
at a given magnetization $m=M/N$ with $M=\sum_{i}S_{i}^{z}$,
we set the magnetic field $h$ to be
the midpoint of the magnetization plateau of $m$.
We employ a dynamical DMRG method~\cite{Jeckelmann2002},
in which we need three target states, i.e.,
$\vert\psi_{\rm G}\rangle$,
$S_{q}^{\alpha}\vert\psi_{\rm G}\rangle$,
and
$[\omega+E_{\rm G}-H+{\rm i}\eta]^{-1} S_{q}^{\alpha} \vert\psi_{\rm G}\rangle$.
We perform dynamical DMRG calculations
with 32 sites at $m=0$ and $m=0.25$ for typical values of $\Delta$.

%-----------------------------------------------------------
% Numerical results
%-----------------------------------------------------------
\section{Numerical results}

%%%%%%%%%%%%%%%%%%%%%%%%%%%%%%%%%%%%%%%%
% Fig. 1
%%%%%%%%%%%%%%%%%%%%%%%%%%%%%%%%%%%%%%%%
\begin{figure}[t]
\begin{center}
\includegraphics[scale=0.7]{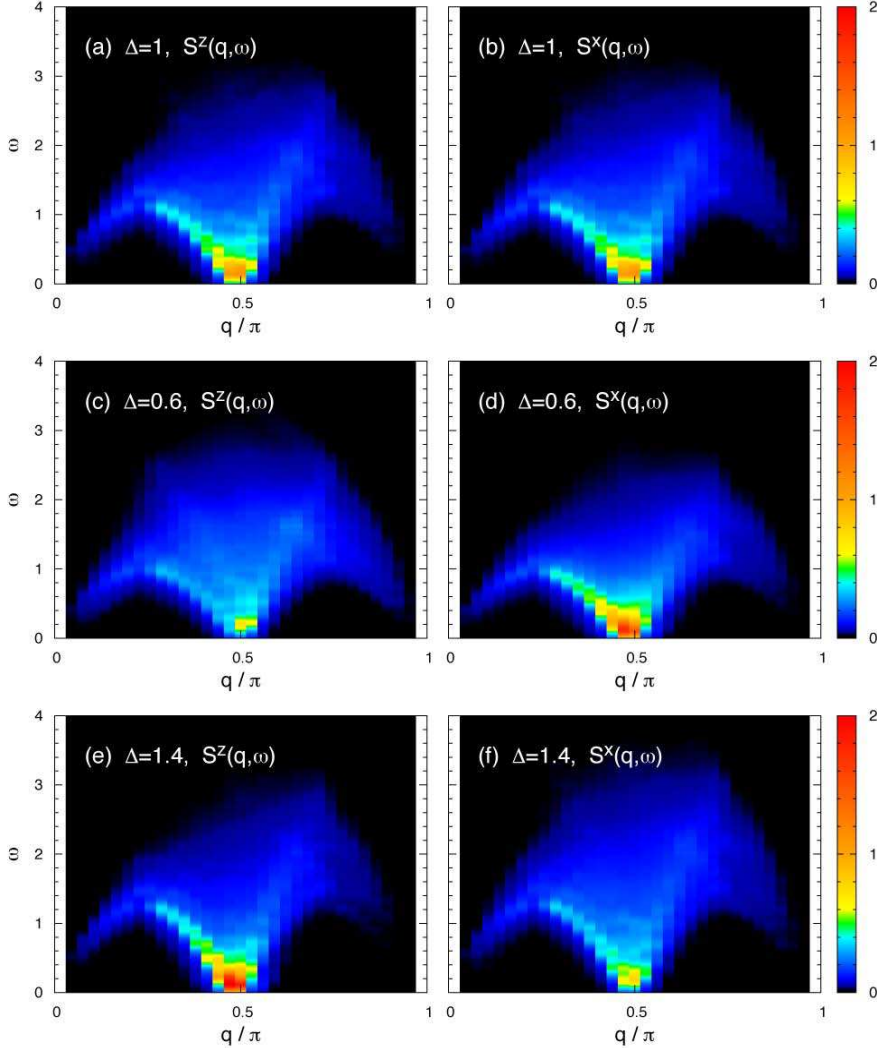}
\end{center}
\caption{
The dynamical spin structure factors $S^{z}(q,\omega)$ and $S^{x}(q,\omega)$
at $J_{1}=-1$, $J_{2}=1$, and $m=0$
for typical values of $\Delta$:
(a),(b) $\Delta=1$ for the isotropic Heisenberg case;
(c), (d) $\Delta=0.6$ for the XY-like case;
and
(e), (f) $\Delta=1.4$ for the Ising-like case.
Three panels in the left column [(a), (c), (e)] are $S^{z}(q,\omega)$,
and
those in the right column [(b), (d), (f)] are $S^{x}(q,\omega)$.
Here, we set $h=0$,
and the system size is $N=32$.
}
\label{fig1}
\end{figure}
%%%%%%%%%%%%%%%%%%%%%%%%%%%%%%%%%%%%%%%%

%%%%%%%%%%%%%%%%%%%%%%%%%%%%%%%%%%%%%%%%
% Fig. 2
%%%%%%%%%%%%%%%%%%%%%%%%%%%%%%%%%%%%%%%%
\begin{figure}[t]
\begin{center}
\includegraphics[scale=0.7]{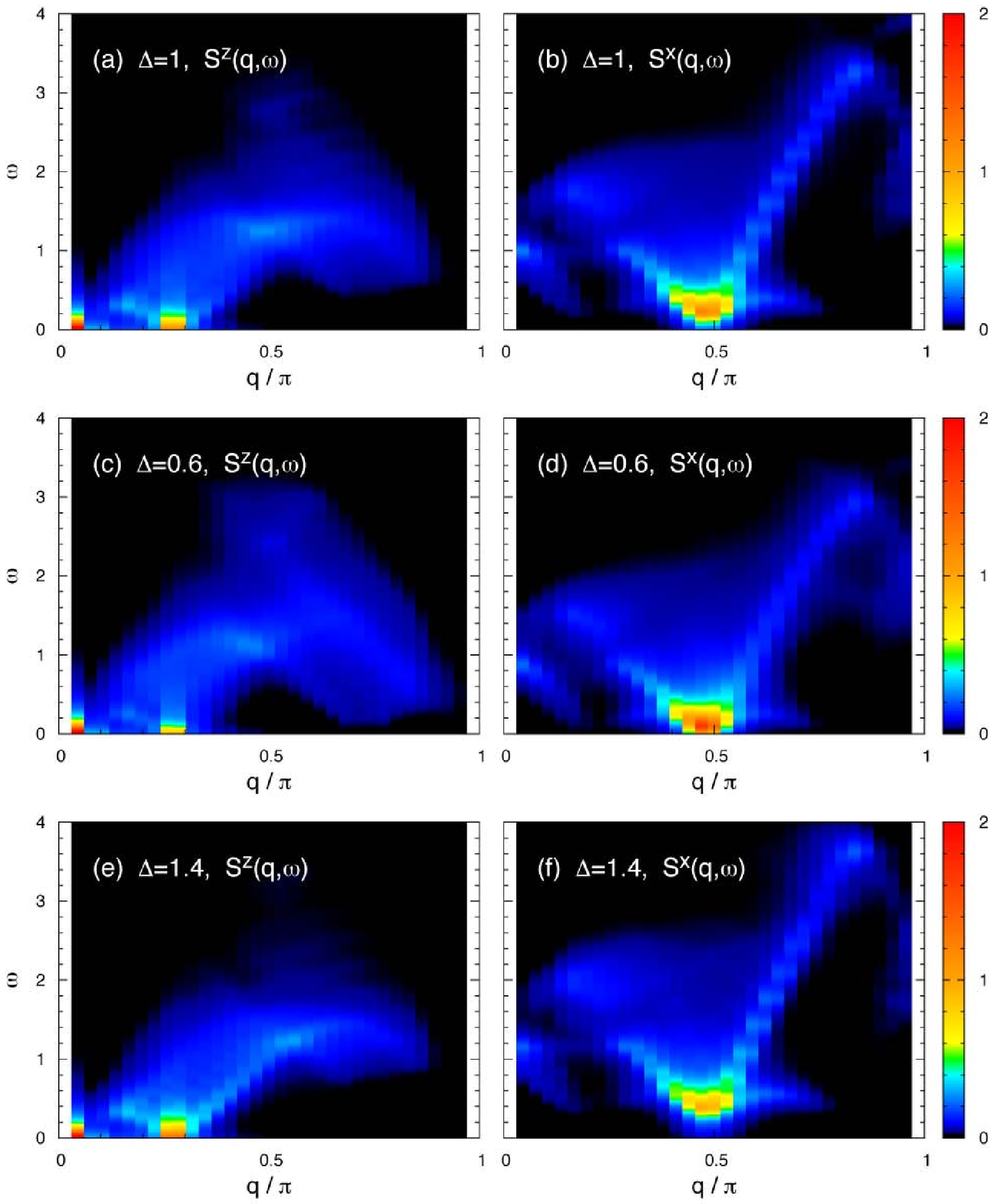}
\end{center}
\caption{
The dynamical spin structure factors $S^{z}(q,\omega)$ and $S^{x}(q,\omega)$
at $J_{1}=-1$, $J_{2}=1$, and $m=0.25$
for typical values of $\Delta$:
(a),(b) $\Delta=1$ for the isotropic Heisenberg case;
(c), (d) $\Delta=0.6$ for the XY-like case;
and
(e), (f) $\Delta=1.4$ for the Ising-like case.
Three panels in the left column [(a), (c), (e)] are $S^{z}(q,\omega)$,
and
those in the right column [(b), (d), (f)] are $S^{x}(q,\omega)$.
Here, we set $h=1.023$ for $\Delta=1$,
$h=0.887$ for $\Delta=0.6$,
and
$h=1.177$ for $\Delta=1.4$,
and the system size is $N=32$.
}
\label{fig2}
\end{figure}
%%%%%%%%%%%%%%%%%%%%%%%%%%%%%%%%%%%%%%%%

First, let us look at spin excitation spectra at zero magnetic field.
In Fig.~\ref{fig1},
we present intensity plots of $S^{z}(q,\omega)$ and $S^{x}(q,\omega)$ at $m=0$.
Figure~\ref{fig1}(a) shows $S^{z}(q,\omega)$ at $\Delta=1$
for the isotropic Heisenberg case,
where the system is in a Haldane dimer phase~\cite{Furukawa2012}.
Here, $S^{z}(q,\omega)=S^{x}(q,\omega)$ due to the SU(2) rotational invariance,
as shown in Figs.~\ref{fig1}(a) and \ref{fig1}(b).
There is a sharp peak at $(q_{0},\omega_{0})=(16\pi/33,0.14)$,
and a large amount of spectral weight is concentrated on this peak.
We note that $\omega_{0}$ seems to be a finite value
because of the finite-size effect.
In fact, $\omega_{0}$ becomes small as the system size increases,
and it is extrapolated to almost zero in the thermodynamic limit.
This is consistent with the renormalization group study
suggesting an exponentially small gap~\cite{Itoi2001}.
The lower boundary of a continuum is found to form a sinusoidal dispersion,
which corresponds to the des Cloizeaux-Pearson mode of spinon excitations
in the limit of $J_{1}/J_{2}=0$,
i.e., decoupled antiferromagnetic chains~\cite{Cloizeaux1962}.
The sinusoidal dispersion is clearly visible due to the large intensity for $q<q_{0}$,
while that for $q>q_{0}$ is less distinct,
since the spectral weight is distributed to a high energy region.
Thus the distribution of the spectral weight is asymmetric
with respect to $q_{0}$~\cite{Enderle2010,Ren2012}.
Note that the intensity extends to high energies
beyond the upper boundary of the two-spinon continuum for $q>q_{0}$,
which is indicative of a four-spinon continuum
suggested in LiCuVO$_{4}$~\cite{Enderle2010}.

In Figs.~\ref{fig1}(c) and \ref{fig1}(d),
we show $S^{z}(q,\omega)$ and $S^{x}(q,\omega)$, respectively,
at $\Delta=0.6$ for the XY-like case,
where the system is in a gapless chiral phase~\cite{Furukawa2012}.
Both $S^{z}(q,\omega)$ and $S^{x}(q,\omega)$ resemble
those in the isotropic Heisenberg case in their overall spectral shapes,
while they move to lower energy.
We observe that
the intensity at the lowest-energy peak is suppressed for $S^{z}(q,\omega)$,
while it is enhanced for $S^{x}(q,\omega)$.
Note that $S^{z}(q,\omega)$ and $S^{x}(q,\omega)$ have the same total intensity,
and the difference of the distribution of the spectral weight
in the $(q,\omega)$ space is essential.
The spin excitation spectra are supposed to be gapless
in the gapless chiral phase,
but we do not clearly see the gapless nature
in the present results because of the finite-size effect.
In Figs.~\ref{fig1}(e) and \ref{fig1}(f),
we show $S^{z}(q,\omega)$ and $S^{x}(q,\omega)$, respectively,
at $\Delta=1.4$ for the Ising-like case,
where the system is in the so-called uudd phase~\cite{Igarashi1989,Tonegawa1990}.
The overall spectral shapes are again similar to
those of the isotropic Heisenberg case,
while they are pushed up to higher energy
in contrast to the XY-like case.
In addition,
the intensity at the lowest-energy peak is enhanced for $S^{z}(q,\omega)$,
while it is suppressed for $S^{x}(q,\omega)$,
which is the reverse of what we find in the XY-like case.
In the uudd phase,
the longitudinal spin excitation spectrum is gapless,
while the transverse one is gapped,
but these features are not clear because of the finite-size effect.
The detailed finite-size scaling analysis is an important future problem,
which will be reported elsewhere.

Now, we move on to the investigation in the magnetic field.
In Fig.~\ref{fig2},
we present intensity plots of $S^{z}(q,\omega)$ and $S^{x}(q,\omega)$ at $m=0.25$,
corresponding to the field-induced SDW regime.
Figure~\ref{fig2}(a) shows $S^{z}(q,\omega)$ at $\Delta=1$
for the isotropic Heisenberg case.
We find a lowest-energy peak at $(q_{0},\omega_{0})=(9\pi/33,0.00)$.
$q_{0}$ moves toward small momentum from the position at zero magnetic field.
By analyzing $q_{0}$ for various values of $m$,
we have confirmed that $q_{0}$ changes with $m$ following a relation
$q_{0}=(1/2-m)\pi$ in the field-induced SDW/SN phase~\cite{Onishi2014},
which supports the bosonization result~\cite{Sato2009}.
On the other hand,
$\omega_{0}$ is nearly zero,
indicating a gapless mode of the longitudinal spin excitation.
Note that we also find a ferromagnetic peak at the origin
due to the finite magnetization.
In Fig.~\ref{fig2}(b), we show $S^{x}(q,\omega)$ at $\Delta=1$.
We find a lowest-energy peak at $(q_{0},\omega_{0})=(16\pi/33,0.22)$.
In particular, we have verified that
$\omega_{0}$ is extrapolated to a finite excitation energy in the thermodynamic limit,
indicating a gapped mode of the transverse spin excitation.
We notice that $S^{x}(q,\omega)=[S^{+}(q,\omega)+S^{-}(q,\omega)]/4$,
i.e.,
it represents the spin excitation changing the magnetization by one.
Since a finite energy is required to break a bound magnon pair,
we have an excitation gap corresponding to the binding energy.

All these features found in the isotropic Heisenberg case
are similarly observed for both
the XY-like [Figs.~\ref{fig2}(c) and \ref{fig2}(d)]
and Ising-like [Figs.~\ref{fig2}(e) and \ref{fig2}(f)] cases.
That is,
for $S^{z}(q,\omega)$,
$q_{0}$ changes with $m$ as $q_{0}=(1/2-m)\pi$ in the field-induced SDW/SN phase,
and $\omega_{0}$ is nearly zero, indicating a gapless mode.
For $S^{x}(q,\omega)$,
$\omega_{0}$ is found at a finite energy,
while it is extrapolated to a finite excitation energy in the thermodynamic limit,
indicating a gapped mode.
Here, we clearly see the difference of $\omega_{0}$ of $S^{x}(q,\omega)$
between the XY-like and Ising-like cases.
Compared to the isotropic Heisenberg case [Fig.~\ref{fig2}(b)],
$\omega_{0}$ decreases in the XY-like case [Fig.~\ref{fig2}(d)],
while it increases in the Ising-like case [Fig.~\ref{fig2}(f)].
Thus the excitation gap is suppressed in the XY-like case,
and it is enhanced in the Ising-like case.

%%%%%%%%%%%%%%%%%%%%%%%%%%%%%%%%%%%%%%%%
% Fig. 3
%%%%%%%%%%%%%%%%%%%%%%%%%%%%%%%%%%%%%%%%
\begin{figure}[t]
\begin{center}
\includegraphics[scale=0.6]{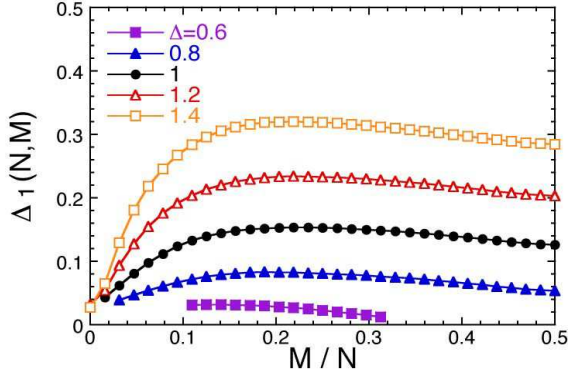}
\end{center}
\caption{
The spin excitation energy $\Delta_{1}(N,M)$
at $J_{1}=-1$ and $J_{2}=1$ for several values of $\Delta$.
Here, we show the results only in the field-induced SDW/SN regime.
$\Delta_{1}(N,M)$ in the VC phase is supposed to be zero
in the thermodynamic limit.
The system size is $N=128$.
}
\label{fig3}
\end{figure}
%%%%%%%%%%%%%%%%%%%%%%%%%%%%%%%%%%%%%%%%

To clarify the dependence of the excitation gap of the transverse spin excitation,
we calculate a spin excitation energy, defined by
\begin{equation}
  \Delta_{1}(N,M) = [E_{0}(N,M+1)+E_{0}(N,M-1)-2E_{0}(N,M)]/2,
\end{equation}
where $E_{0}(N,M)$ is the lowest energy
of the $N$-site system in the subspace of $M$ at $h=0$.
Here, we use an ordinary DMRG method, not dynamical one, to obtain $\Delta_{1}(N,M)$.
In Fig.~\ref{fig3}, we plot $\Delta_{1}(N,M)$
for several values of $\Delta$.
The results in the field-induced SDW/SN regime are shown.
Note that we find reentrant behavior in the XY-like case,
i.e., the ground state changes from the VC state to the SDW/SN state and turns to the VC state again
as the magnetization increases from zero to the saturation~\cite{Heidrich-Meisner2009}.
We find that $\Delta_{1}(N,M)$ increases as $\Delta$ increases
over the XY-like and Ising-like regions.
This indicates that the easy-plane anisotropy disfavors the formation of two-magnon bound states,
and the easy-axis anisotropy stabilizes the two-magnon bound state,
which is consistent with the behavior of correlation functions~\cite{Heidrich-Meisner2009}.
%The precise determination of the excitation gap in the thermodynamic limit
%is an important future problem.

%-----------------------------------------------------------
% Summary
%-----------------------------------------------------------
\section{Summary}

We have investigated the spin excitation dynamics of
the spin-1/2 one-dimensional $J_{1}$-$J_{2}$ XXZ model in the magnetic field
by numerical methods.
We have observed that in the field-induced SDW/SN phase,
the longitudinal spin excitation is gapless,
while the transverse spin excitation is gapped,
irrespective of the exchange anisotropy.
The excitation gap of the transverse spin excitation
increases as the exchange anisotropy increases,
which is indicative that the easy-axis anisotropy stabilizes the two-magnon bound state.

%-----------------------------------------------------------
% Summary
%-----------------------------------------------------------
\ack

The author thanks
T. Masuda, S. Maekawa, M. Mori, and T. Sugimoto for useful discussions.
Numerical calculations have been supported by
the supercomputer centers at the Japan Atomic Energy Agency
and the Institute for Solid State Physics, the University of Tokyo.

%-----------------------------------------------------------
% Reference
%-----------------------------------------------------------

\end{document}